\documentclass[aps,pre,twocolumn,superscriptaddress,floats,showpacs]{revtex4-2}
\usepackage[utf8]{inputenc}
\usepackage[english]{babel}
\usepackage{color}
\usepackage{graphicx}
\usepackage{hyperref}
\usepackage{amsmath}
\usepackage{textcomp}
\usepackage{setspace}
\usepackage{geometry}
\begin{document}

\title{\bf Compression of high-power laser pulse leads to increase of electron acceleration efficiency}
\author{O. E. Vais}
\affiliation{P. N. Lebedev Physics Institute,
	Russian Academy of Science, Leninskii Prospect 53, Moscow 119991,
	Russia}
\affiliation{Center for Fundamental and Applied Research,
	Dukhov Research Institute of Automatics (VNIIA), Moscow 127055, Russia}
\author{M. G. Lobok}
\affiliation{P. N. Lebedev Physics Institute,
	Russian Academy of Science, Leninskii Prospect 53, Moscow 119991,
	Russia}
\affiliation{Center for Fundamental and Applied Research,
	Dukhov Research Institute of Automatics (VNIIA), Moscow 127055, Russia}
\author{V. Yu. Bychenkov}
\affiliation{P. N. Lebedev Physics Institute,
	Russian Academy of Science, Leninskii Prospect 53, Moscow 119991,
	Russia}
\affiliation{Center for Fundamental and Applied Research,
Dukhov Research Institute of Automatics (VNIIA), Moscow 127055, Russia}

%\date{}

\begin{abstract}
Propagation of ultrarelativistically intense laser pulse in a self-trapping mode in a near critical density plasma makes it possible to produce electron bunches of extreme parameters appropriate for different state of art applications. Based on the 3D PIC simulations, it has been demonstrated how the best efficiency of electron acceleration in terms of the total charge of high-energy electrons and laser-to-electrons conversion rate can be achieved. For given laser pulse energy the universal way is a proper matching of laser hot spot size and electron plasma density to the laser pulse duration. The recommendation to achieve the highest yield of high-energy electrons is to compress laser pulse as much as possible. As example, compression of the few tens fs pulse to the $\sim 10$ fs pulse leads to generation of the high-energy electron bunch with the highest total charge to exhibit conversion efficiency exceeding $50\%$ for the Joule-level laser pulse energies.
\end{abstract}

\maketitle
		
\section{Introduction}{\label{sec1}}
Laser-plasma accelerators are of high interest due to their extremely large accelerating field gradients that allows receiving high-energy electron beams of application demand on a scale of hundred microns. Depending on laser-plasma parameters, there are a number of underlying mechanisms, which lead to different space-energy distributions of the accelerated particles. Ones, such as the laser wakefield accelerator (LWFA), the
plasma beat wave accelerator (PBWA), the self-modulated LWFA and so on, consist in the electron acceleration by the longitudinal field of plasma waves exciting by laser pulses in low density plasma \cite{Esarey_2009}. In the highly nonlinear regime, a solitary laser-plasma structure, "bubble'' \cite{Pukhov_2002}, is formed allowing, in particular, the use of a higher plasma density. In a rare plasma, spectra of accelerated electrons have the monoenergetic feature with the peak at the highest energies up to several GeV \cite{Wang_2013, Clayton_2010}. The total charge of such particles is typically at the pC level. Another mechanism, named as direct laser acceleration (DLA), is associated with the acceleration of electrons being in the betatron resonance with the laser frequency \cite{Pukhov_1999, Gahn_1999, Mangles_2005}. It appears when a picosecond high-energy laser pulse propagates in a denser plasma by forming a plasma channel. In this case, the high-energy part of an electron spectrum has the exponential form but demonstrates high total charge at the level of a few microcoulombs \cite{Rosmej_2020}. Although the spectra of electrons accelerated in the "bubble" and DLA regimes are different, both of them are very effective showing that the conversion rate of the laser energy to the high-energy electron can be as high as $\approx 20\%$ \cite{Gordienko_2005, Rosmej_2020}.

For the ultrarelativistic laser intensities, the recent studies identified and substantiated a stable regime of laser pulse propagation and electron acceleration in the near critical density plasma \cite{Bychenkov_2019, Lobok_2019}. The balance of diffraction divergence and relativistic nonlinearity (relativistic mass increase and cavitation) in a plasma provides the soliton laser-plasma structure \cite{Kovalev_2020} -- "laser bullet",  which keeps its shape with only weak swelling for many Rayleigh lengths. This regime is physically similar to self-trapping of weak electromagnetic waves, which is described by the Schr\"{o}dinger equation with a cubic nonlinearity \cite{Talanov_1964, Chiao_1964, Ahmanov_1966}, that is why this regime was named as a relativistic self-trapping regime (RST) \cite{Bychenkov_2019}. The laser pulse forms the plasma cavity, which is fully filled by the trapped light. The RST regime in the form of "laser bullet" complements the previously observed "bubble" regime of RST \cite{Pukhov_2002}, which occurs for the pulse length, $L$, shorter than its transversal size, $D$. Thus, the "bubble" and "laser bullet" regimes clearly appear under two different characteristic conditions, when the pulse duration is shorter than the cavity collapse time ($D/c$), i.e. $D\gg L$, or comparable to the latter, $L\simeq D$, correspondingly.   

Accelerated particles are affected by the laser pulse field as well as by the plasma cavity field in the RST "laser bullet" regime. It has been shown \cite{Lobok_2019, Lobok_2021}, that the laser pulse affects an injection of electrons and an angular particle distribution, while electrostatic cavity field mainly contributes to electron energy gain. The RST regime results in a substantial proportion of high-energy particles, that for near critical density plasma causes record total charge of the generated electron bunch as compared to other acceleration mechanisms by lasers of the same energy. The advantage of such high electron charge has already been used to predict highly efficient bright synchrotron radiation \cite{Lobok_2021}. Here, the further development of the control methods of RST mode is presented with the aim of its optimization by a pulse duration.
  
Previously, we have studied the RST regime for the 30 fs duration high-power laser pulses, in excess of 100 TW and demonstrated the record charge for the generated sub-GeV electron bunches at multi-nC level \cite{Bychenkov_2019}. At the same time, the question to which extent a laser pulse shortening may improve electron characteristics is still open. Moreover, this question has become very relevant in the light of recent achievements in shortening high-power laser pulses. State-of-the-art technology makes it possible shortening the multi-Joule energy femtosecond laser pulses almost without loss of energy with so-called CafCA (compression after compressor approach) \cite{Khazanov_2019,Ginzburg_2020}. We have already discussed how pulse shortening affects RST regime for moderate (multi-TW) laser pulse power \cite{our_JETPhLetters}. At the same time, this question requires broader consideration over a wider range of laser energies, in particular, to cover the most interesting for applications laser-plasma parameters. Here we consider ultra-relativistic case, $a_0\gg 1$, and compare electron acceleration with the basic pulse (40~fs) and that after his 4-fold shortening up to PW-level laser pulse power. For the long (40~fs) laser pulse, we consider different regimes of its propagation to analyze in details the each of them features to make sure no way to have advantage over the shortest pulse. Our aim is to quantify by using 3D PIC simulations how the total charge of high-energy electrons, conversion efficiency and characteristic particle energy change (1) with pulse duration at the given laser energy and (2) with laser pulse energy at different pulse duration. 

This paper is organized as follows. Section~\ref{sec2} summarizes rough estimates for the characteristics of the laser-plasma structure and the electron bunch accelerated. In sections \ref{sec3} and \ref{sec4}, we consider the simulations of the RST regime of the laser propagation and the laser self-modulation, respectively. Then we discuss the energy spectra of accelerated electrons and corresponding characteristics of electron bunch, which were obtained in previous sections. After that, in section~\ref{sec6} we analyze the results for another laser energies to generalize our findings to a wider range of laser parameters.   

\section{Rough directions}{\label{sec2}}

\begin{figure}
\center{\resizebox{0.55\hsize}{!}{\includegraphics{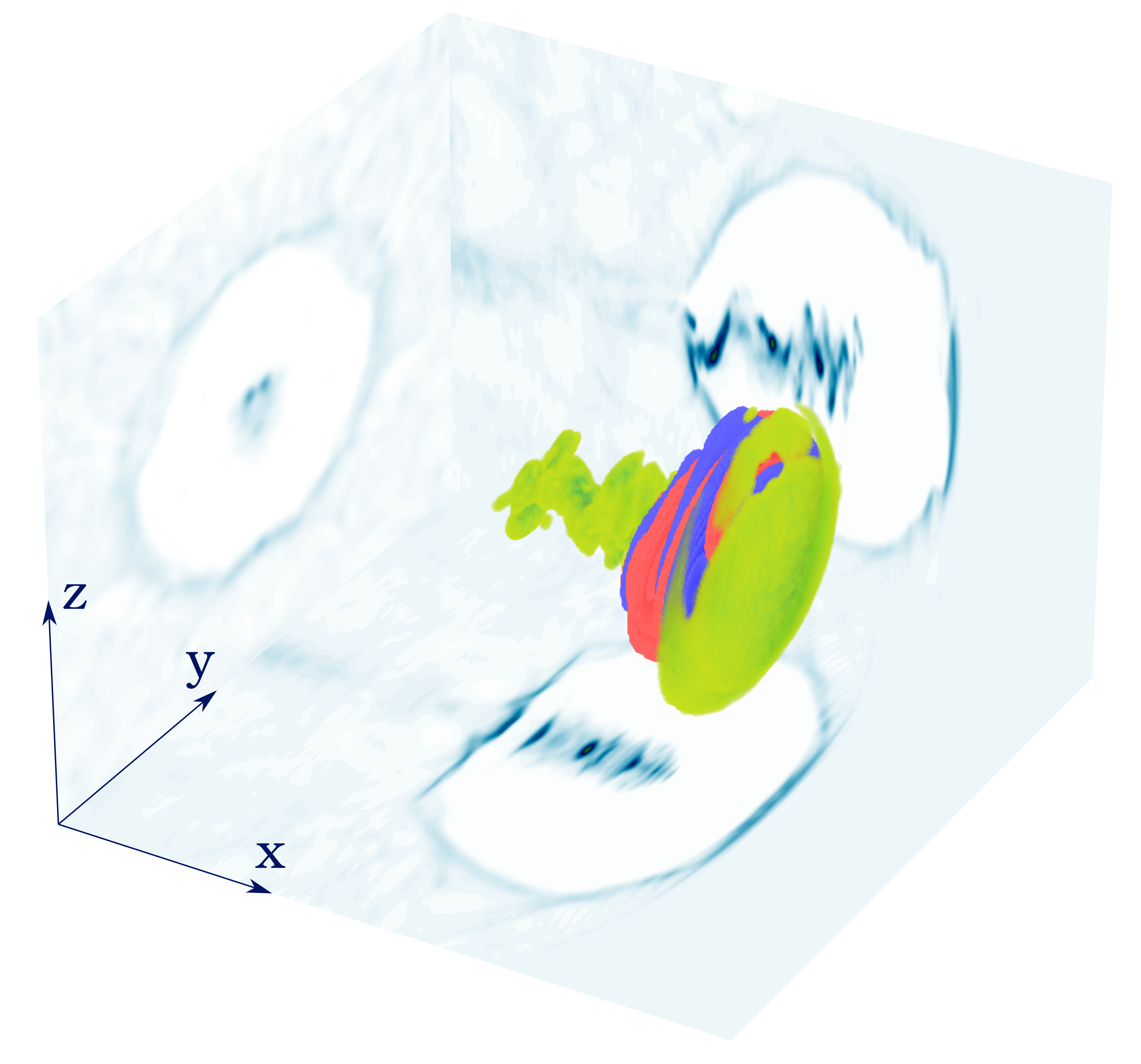}}}
\caption{The 3D distributions of the electron density (light green) and the laser field amplitude (blue-red) from PIC simulation. Side panels display central slices of the plasma cavity in white-blue, namely the xz-plane is the laser polarization plane, the yz-plane is perpendicular to the direction of the laser propagation and the xy-plane is perpendicular to the listed ones.}
\label{ris:scheme}
\end{figure}

It has been shown before \cite{Bychenkov_2019, Lobok_2019, Kovalev_2020} that for the relativistic laser intensities the RST-regime is able to provide stable laser pulse propagation in a near-critical density plasma over many Rayleigh lengths in the form of plasma cavity filled by a laser field (the illustration is Fig. \ref{ris:scheme}). As a result, a maximum total charge of high-energy electron bunch and, correspondingly, conversion efficiency are achieved. This occurs in the matching condition for the cavity diameter $D$, the electron plasma density $n_e$, and the laser field amplitude,
\begin{equation} 
    D \approx \lambda_p \simeq 2.6 \frac{c}{\omega_l}\sqrt{a_0\frac{n_c}{n_e}} \,, \quad  a_0\gg 1\,, \label{RST}
\end{equation}
where $\omega_l$ and $\omega_p$ are the laser light and electron plasma frequencies, respectively, $\lambda_p=(2\pi c/\omega_p)\sqrt{\gamma}$ is the plasma wavelength, $\gamma=\sqrt{1+a_0^2/2}\simeq a_0/\sqrt{2}$ is the electron relativistic factor, $a_0=e E_L/m_e\omega_l c$ is the standard dimensionless laser field amplitude ($E_L$), $e$ and $m_e$ are the electron charge and mass, and $n_c$ is the critical electron plasma density. Cavity diameter is of the order of a laser focal spot and slowly evolves as it propagates. To ensure a stable pulse evolution to the RST-regime the laser focal spot diameter, $D_L$, should be somewhat less than the steady-state one Eq. ({\ref{RST}}) \cite{Lobok_2018}.   

The condition Eq.~(\ref{RST}) has been deduced from the qualitative geometric-optical treatment of RST \cite{Lobok_2019} and nonlinear Schr\"{o}dinger equation approach \cite{Kovalev_2020}. 
As demonstrated in Ref. \cite{Kovalev_2020}, Eq.~(\ref{RST}) can be derived from \textit{ad hoc} replacement of the electron mass to the relativistic electron mass in the weak-field analogue of the matching condition for the laser beam self-trapping in the nonlinear medium with cubic nonlinearity corresponding to ${a_0}\ll1$ \cite {Talanov_1964, Chiao_1964, Ahmanov_1966}. The requirement for the matched transversal cavity size Eq.~(\ref{RST}) also follows from the balance of the radial ponderomotive force of the laser pulse and the Coulomb force of the ion channel \cite{Sun_1987, Kostyukov_2004, Lu_2006}. On the other hand, numerical simulations confirm this matching with only some change in the numeric factor in Eq. (\ref{RST}) due to different initial conditions for initiation of RST, i.e. $D \simeq 2.24c\sqrt{a_0n_c/n_e}/{\omega_l}$ \cite{Gordienko_2005}, $D \simeq 4c\sqrt{a_0n_c/n_e}/{\omega_l}$ \cite{Lu_2007}, $2c\sqrt{a_0n_c/n_e}/{\omega_l}$  \cite{Poder_2024}, etc.  For short laser pulse, with the length considerably shorter than the focal spot size the RST cavity is the electron-empty sphere with a laser "snow plow" ahead. This RST-regime was first observed in the 3D PIC simulations \cite{Pukhov_2002} and later widely referenced as the “bubble” regime. The study \cite{Gordienko_2005} showed that the number of electrons accelerated in this mode is proportional to ${a_0}$, that is also in the discussed “laser bullet” case (see, Eq. (\ref{density})). 
	
In the context of stable propagation of a laser pulse, it is important to take into account that when its longitudinal or transverse size exceeds the relativistic plasma wavelength the pulse is subject to such instabilities as self-modulation or filamentation, correspondingly, \cite{Esarey_2009}. The condition Eq. (\ref{RST}) prevents filamentation of the pulse and to provide a longitudinal stability the pulse length, $L=c\tau$, should be limited as follows 
\begin{equation}\label{mod}
	L \lesssim \lambda_p, \quad  \text{ or } \,\,  \sqrt{n_e/n_c} \lesssim 5 \sqrt{a_0}/\omega_l \tau\,,
\end{equation}    
where $\tau$ is the laser pulse duration, $\lambda_p$ is the classical plasma wavelength. At the same time, the denser the plasma, the greater the total charge of accelerated electrons. Because of that, we accept in Eq. (\ref{mod}) the upper limit for the electron density, 
exclude the inequality in Eq. (\ref{mod}) and use
\begin{equation}\label{density}
	n_e/n_c \approx 25 a_0/( \omega_l \tau)^2\,.
\end{equation}    
This choice of density leads to approximately equality of the laser pulse length to the laser-plasma cavity diameter,  i.e. Eq. (\ref{RST}) reads
\begin{equation}\label{R_tau}
c\tau \approx D\,.
\end{equation} 
Below we consider the order of magnitude acceleration characteristics scalings make it possible from rough parametric estimates.

While the laser beam waist size reduction can be achieved with larger apertures, the laser pulse duration can be diminished by spectrum broadening in nonlinear crystals \cite{Khazanov_2019}. So, for the laser pulses with the same energy, $W_L$, it's worth to consider different pulse durations and the case of near-spherical cavity with the aim to reach most efficient electron acceleration under the condition
\begin{equation}\label{W_L}
W_L \sim a_0^2n_c m_ec^5\tau^3=const\,.
\end{equation} 
Then, for this optimum case of the near spherical laser pulses,  Eqs. (\ref{density}) -- (\ref{W_L}) result to the following scaling
\begin{equation}\label{tau}
n_e \propto \sqrt{W_L/\tau^7}\,,
\end{equation}
which shows rather sharp density increase with pulse duration decrease.

The spectrum, electron bunch total charge and average and maximum particle energies are usually considered as indicative characteristics, which demonstrate an efficiency and interest to applications of the considered laser-plasma accelerator. It is natural to assume that similar to Refs. \cite{Katsouleas_1987, Gordienko_2005, Jansen_2014} the total charge of the accelerated particles, $Q_0$, in the RST regime is proportional to the number of self-injected into the cavity electrons, which, in turn, is expected to be proportional to the cavity Coulomb charge, i.e.
\begin{equation}\label{Q_0}
	Q_0 \sim e n_e D^3 \propto a_0D \propto \sqrt{W_L/\tau}\,.
\end{equation}
Similar scaling of the number of accelerated electrons derived for $a_0 \gg 1$ has also been presented in Ref. \cite{Gordienko_2005} for the quasimonoenergetic electrons in the bubble. 

The characteristic electron energy gain, $\varepsilon_e$, is given by $eEl_{acc}$, where $E \sim (m_ec^2/e)\times (\omega_p/c)^2 D \propto n_e D$ is the maximum plasma cavity Coulomb field and $l_{acc}$ is the acceleration length \cite{Lobok_2018,Decker_1996, Lu_2007}. In the RST regime, when the laser propagation distance far exceeds the Rayleigh length the value $l_{acc}$ is determined by the lesser of dephasing length, $l_{dph}$, and pump depletion length, $l_{dpl}$. The dephasing length corresponds to the distance which trapped electrons travel until they enter the decelerating cavity field phase and the pump depletion length is the distance, over which a laser pulse losses its energy as result of electron raking by light pressure and producing the cavity electrostatic field. In the case of modulation-stable laser pulse propagation, when the condition Eq. (\ref{mod}) is satisfied, one gets $l_{dpl} \sim l_{dph}$, i.e. the acceleration distance is $l_{acc} \sim l_{dpl}$, where \cite{Lu_2007, Gordienko_2005}: 
\begin{equation}\label{depletion}
	l_{dpl} \propto a_0 c\tau(n_c/n_e)\,.
\end{equation}
Correspondingly, the electron characteristic energy scaling can be presented in the form:
\begin{equation}\label{W_max}
	\varepsilon_e \propto a_0 D \tau \propto (W_L \tau)^{1/2}\,.
\end{equation}
These scaling can also be addressed to the "bubble" regime of a wide laser pulse, $c\tau \ll D$, where only small front part of a cavity is filled with laser light during entire propagation distance (cf. the formula (1) from Ref. \cite{Gordienko_2005}). Note, that for the optimum near spherical laser bullet the last two scalings transform to 
\begin{equation}\label{depletionS}
	l_{dpl} \propto D^3
\end{equation}
and 
\begin{equation}\label{W_maxS}
	\varepsilon_e \propto (W_L D)^{1/2}\,,
\end{equation}
correspondingly.

Applying these qualitative estimates and scalings to a laser pulse with given energy, one can conclude from (\ref{Q_0}) that shorter pulse duration is preferred to reach higher electron bunch charge in the case of near spherical laser bullet. As concerning conversion efficiency of the laser-to-electrons energy, $\eta$, its estimate can be insufficient because of very simplified guess $\eta \propto Q_0\times \varepsilon_e \propto W_L$, which does not show $\tau$-dependence and also because of ignoring a possible impact of the pulse duration on the electron spectrum shape. For this reason, we performed the 3D PIC simulations, the results of which are presented below to quantify introduced electron characteristics and find to which extent the above rough directions can work.

\begin{figure*} 
\begin{minipage}{\linewidth}
 \center{\resizebox{0.85\hsize}{!}{\includegraphics{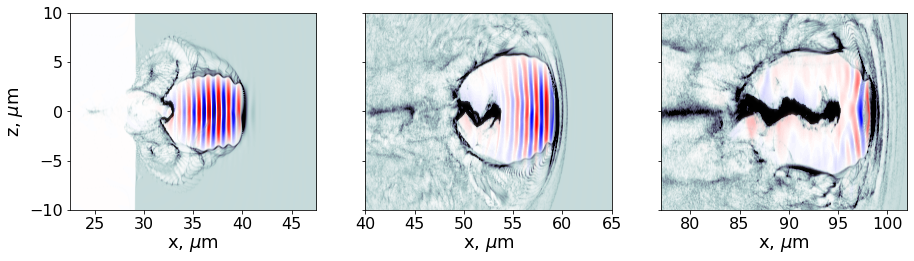}}}
\end{minipage}
\vfill
\begin{minipage}{\linewidth}
 \center{\resizebox{0.85\hsize}{!}{\includegraphics{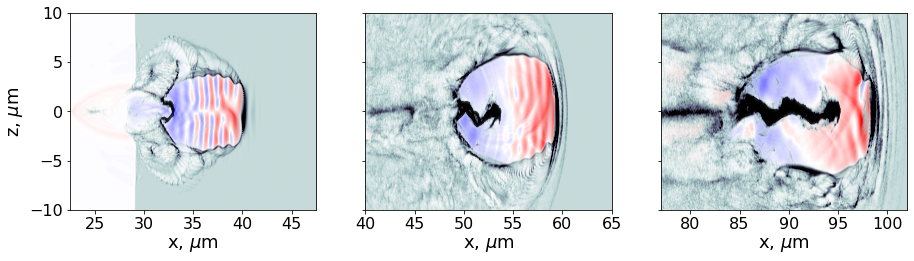}}}
\end{minipage}
\caption{Dynamics of the plasma cavity during laser pulse ($\tau=10$~fs) propagation trough a plasma with $n_e = 0.15 n_c$. Electron density is shown in blue-gray. Sequential image frames show the laser pulse $E_z$-component (top) and the $E_x$ longitudinal electric field (bottom) in blue-red.}
\label{ris:prop10}
\end{figure*}

\begin{figure*}
\begin{minipage}{\linewidth}
 \center{\resizebox{0.85\hsize}{!}{\includegraphics{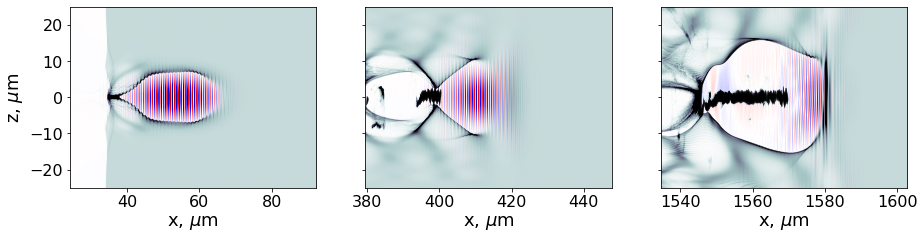}}}
\end{minipage}
\vfill
\begin{minipage}{\linewidth}
 \center{\resizebox{0.85\hsize}{!}{\includegraphics{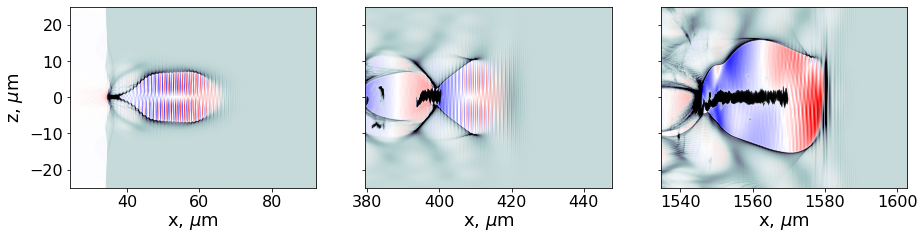}}}
\end{minipage}
\caption{Dynamics of the cavity electric fields for 40~fs laser pulse propagating trough a plasma with $n_e = 0.005 n_c$. Electron density is shown in blue-gray/black. Sequential image frames show the laser pulse $E_z$-component (top) and the $E_x$ longitudinal electric field (bottom) in blue-red.}
\label{ris:prop40low}
\end{figure*}

\section{Simulations of the RST "laser bullet" regime}\label{sec3}

By using the high performance electromagnetic 3D PIC code \cite{VORPAL} we have studied the laser pulse propagation in the RST regime and corresponding electron acceleration for different pulse durations at the given laser energy $W_L \simeq 2.2$~J. The linear polarized laser pulse with the wavelength $\lambda = 1\, \mu$m propagated in the $x$-direction. It was focused on the front side of a fully-ionized homogeneous dense gas plasma target consisting of He-ions and electrons. The plasma model is justified due to the relativistically intense laser pulses considered, which easily ionize the target either by a natural prepulse or by the pulse itself at its very leading edge. To analyze a pulse duration impact on the electron acceleration, the Gaussian laser pulse FWHM duration, $\tau$, was varied in the range of 10 -- 40~fs. Note, that PW laser pulses of $\sim 10$ fs duration are now available with CafCA-shortening for the standard femtosecond multi-Joule lasers \cite{Khazanov_2019,Ginzburg_2020}. The simulations with a moving window used the spatial grid steps $0.02\lambda \times 0.1\lambda \times 0.1\lambda$. The simulation window size was chosen to be adequate to the laser pulse size within a plasma, $X\times Y \times Z = 58\lambda \times 60\lambda \times 60\lambda$.   
 
First, we considered propagation of the 10~fs laser pulse with $a_0 \approx 42$ in a near-critical density plasma, $n_e = 0.15 n_c$. Laser-plasma dynamics is illustrated in Fig.~\ref{ris:prop10} by evolution of the electron density, the $E_z$-component of the laser pulse and the longitudinal electric field $E_x$. The FWHM focal spot, $D_{L}$, was 2.8~$\mu$m, i.e. laser pulse had a spherical form satisfying Eq.~(\ref{R_tau}). At same the time, for $a_0>>1$ it is approximately twice less than the cavity diameter $D$ \cite{Kovalev_2024}. Simulation shows that the cavity diameter quickly sets and then changes only in a quasi-stationary manner, slowly increasing during pulse propagation from $D\simeq 11.5 \lambda$ (Fig. \ref{ris:prop10}, middle frames) to $D\simeq 13 \lambda$ (Fig. \ref{ris:prop10}, end frames).

Simulation demonstrates that conditions Eqs. (\ref{RST}), (\ref{density}), (\ref{R_tau}) are reasonably suited to laser pulse capture in the formed plasma single cavity fully filled by light ("laser bullet").
Such a strongly nonlinear laser-plasma structure propagates through the target over approximately 80~$\mu$m without plasma wave formation. During propagation, the laser pulse continuously depletes due to strong etching at the leading edge and because of that the "laser bullet" regime transforms finally to the "bubble" regime \cite{Pukhov_2002}. This is clearly seen at the top of Fig. \ref{ris:prop10}. Electrons are injected into the cavity at its rear side throughout the entire time. Initially, they are accelerated by the laser field and the longitudinal cavity field, but as the laser pulse shortens, its role in acceleration disappears. The dynamics of the electron bunch (in black) acceleration and the $E_x$-field evolution are illustrated at the bottom of Fig. \ref{ris:prop10}. The bunch of accelerated electrons is modulated in the $xz$-plane. The explanation could be related to the carrier-envelope phase effects (CEP), that could manifest itself in the form of electron bunch modulation \cite{Nerush_2009}.

For the 40~fs laser pulse of the same energy and $a_0 \approx 10$, we chose the maximum plasma density $n_e = 0.005n_c$, for which the laser pulse self-modulation is still absent, and increased the laser focal spot size to 5.5~$\mu$m. However this increase was not enough for the considered electron concentration. The laser beam propagating through the plasma more expanded to 8~$\mu$m with $a_0 \approx 7$, until the diffraction divergence was not balanced by the relativistic self-focusing, i.e. the laser-plasma structure was attracted by the RST-regime.
In the steady state, the plasma cavity diameter was about 24~$\mu$m (Fig.~\ref{ris:prop40low}, middle frames), which also slowly increased to 35~$\mu$m during the pulse propagation (Fig.~\ref{ris:prop40low} right).
Although the laser pulse length is greater than its steady-state diameter, it was not enough to seed laser modulations. It is in accordance with the prediction of the paper \cite{Andreev_1995}, where for self-modulation effect a laser pulse should have a duration of several plasma-wave periods (namely $L/\lambda_p > 3$).    

In the optimum (near-spherical) RST regime for a given laser pulse energy, a depletion length substantially depends on the pulse duration and significantly increases with the latter increase. From the estimate Eq.~(\ref{depletionS}) the depletion length ratio for short (1) and long (2) pulses is $l_{dpl}^{(1)}/l_{dpl}^{(2)} \approx (D^{(1)}/D^{(2)})^{3} \approx 23.3$. Thus the depletion length increases from $\approx 80$~$\mu$m to $\approx 2$~mm that was observed in the PIC-simulations.  

In the case of the 40~fs pulse duration, a plasma wave behind the light-carrying soliton cavity (see Fig.~\ref{ris:prop40low}) is excited in contrast to a single soliton cavity for the 10~fs pulse. Nevertheless, the plasma wave acceleration of the electrons is negligible as compared to the electron acceleration in the first light-carrying cavity.
We attribute a plasma wave excitation for the 40~fs pulse to the not high enough $a_0$, which should significantly exceed 1 ($a_0\gtrsim 10$) for the ideal "laser bullet" regime. In accordance with  Eq. (\ref{W_L}), the requirement for high laser field amplitude limits the pulse duration as follows 
\begin{equation}\label{Rel}
	(c\tau)^3 \ll \frac{W_L}{n_cm_ec^2} \,,
\end{equation}
i.e. the higher pulse energy the longer pulse duration is acceptable for the "laser bullet" RST regime. Nevertheless, for the considered laser energy, the proper laser-plasma matching makes it possible to avoid self-modulation instability for the 40~fs pulse and to provide the self-injection of electrons \cite{Mangles_2012}, that could not be realized for the laser pulses with lower energies 
\cite{Faure_2004, our_JETPhLetters}.  Note also, that in this regime the injected electrons interact with the laser field that leads to particle oscillations in the polarization plane with the period approximately equal to the laser wavelength \cite{Lobok_2018,Nemeth_2008}. 

For given laser power, the easiest way to increase of $a_0$ is to decrease of $D_{L}$, that can be achieved by optics with shorter focal length. In this case the laser length can be noticeably greater than the beam diameter $c\tau > D$. However, this is the case of a laser self-modulation and the question naturally arises whether or not the corresponding mismatched regime is able to lead to enough effective electron acceleration, comparable to that from the "laser bullet" RST regime at the above considered matched condition Eq. (\ref{R_tau}). Another interesting question is whether or not such mismatched regime could evolve to the laser bullet one, since the soliton nature of the latter might behave as an attractor for certain values of laser-plasma parameters. The answers to these questions are in the next section.

\section{Laser self-modulation regime}{\label{sec4}}

\begin{figure*}
\begin{minipage}{\linewidth}
 \center{\resizebox{0.85\hsize}{!}{\includegraphics{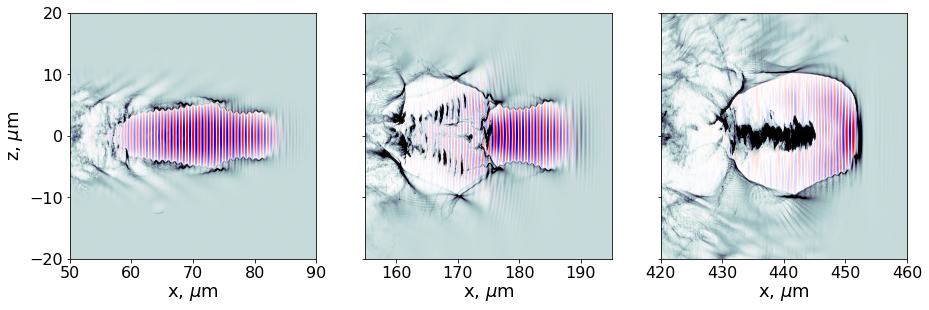}}}
\end{minipage}
\hfill
\begin{minipage}{\linewidth}
 \center{\resizebox{0.85\hsize}{!}{\includegraphics{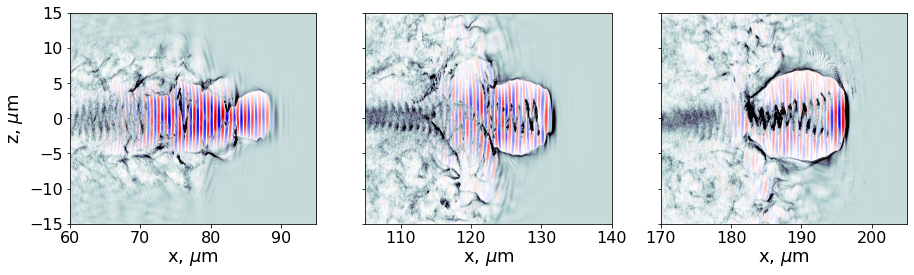}}}
\end{minipage}
\caption{Dynamics of the laser-plasma structure for the 40~fs laser pulse propagating trough the plasma with $n_e = 0.02 n_c$ (top frames) and $n_e = 0.065 n_c$  (bottom frames). Electron density and $E_z$-component of the laser pulse are shown in blue-gray and blue-red, correspondingly.}
\label{ris:prop40_MM}
\end{figure*}

Here, we again consider the 40-fs 2.2~J laser pulses but with increased $a_0$ by tighter focusing. To analyze the nonlinear evolution of the self-modulation process when the condition Eq.~(\ref{R_tau}) is violated, we chose $D_{L} = 4.2~\mu\text{m}$ and  $D_{L}=2.8~\mu\text{m}$. In both cases the normalized laser-filed amplitude exceed 10 (about $14$ and $21$). The densities of a plasma target were taken 0.02$n_c$ and 0.065$n_c$, respectively. Figure~\ref{ris:prop40_MM} shows the dynamics of these laser pulses during their propagation through the plasma. The laser pulse lengths exceeded about 3 and 4 times the beam diameters that leads to the self-modulation of the laser pulse. This is clearly seen in middle frames in Fig.~\ref{ris:prop40_MM}. The self-modulation observed is related to a transverse redistribution of the pulse energy. Such transverse laser energy loss is in accordance with prediction of the linear theory by Andreev et al. \cite{IEEE_TRANS._PLASMA_SCIENCE-1996}. The greater $c\tau/D_L$, the greater transverse energy transport. So, we expect a decrease in the efficiency of electron acceleration compared to the cases discussed in Sec.~\ref{sec3}. 
After the transverse energy release has occurred, the near spherical cavities filled by light (laser bullets) are formed (Fig.~\ref{ris:prop40_MM}).

The laser pulses completely deplete at the distances $l_{dpl}^{(1)}\simeq 750 \mu$m and $l_{dpl}^{(2)}\simeq 340 \mu$m in the plasma target with the densities of 0.02$n_c$ and 0.065$n_c$, respectively. It agrees with the scaling Eq.~(\ref{depletion}) for non-spherical laser pulses. It gives for the depletion length ratio $l_{dpl}^{(1)}/l_{dpl}^{(2)} \simeq (D^{(1)}/D^{(2)})^2 \simeq 2.37$ that well matches the PIC result: $l_{dpl}^{(1)}/l_{dpl}^{(2)}\simeq 750/340$. % 
Simulations show that the main electron acceleration occurs after the laser bullet regime is established. 

We have discussed two regimes of the laser propagation through the plasma target. Both of them was followed by the acceleration of the self-injected electron bunch. To analyze the efficiency of this process, we consider below the electron spectra and their characteristics. Moreover this analysis can answer the question whether shortening the laser pulse leads to the growth of the acceleration efficiency.

\section{Electron bunch characteristics}{\label{sec5}}

\begin{figure*}
\center{\resizebox{0.85\hsize}{!}{\includegraphics{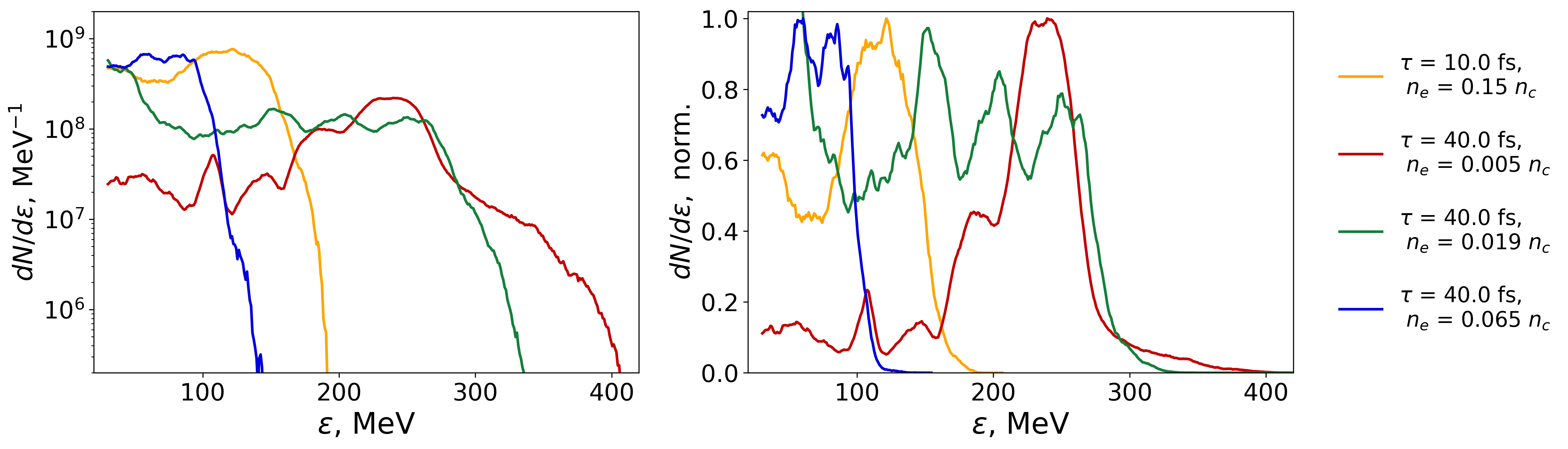}}}
\caption{Electron spectra which are formed at the instant when electron bunch charge reaches the maximum value for different laser pulses in the logarithmic (left) and linear (right) scales.}
\label{ris:sp}
\end{figure*}

In all considered cases (Secs.~\ref{sec4} and \ref{sec5}), electron energy distributions demonstrate a plateau on a logarithmic scale with some quasi-monoenergeticity on a linear scale at the high-energy part of the spectrum near the cutoff as displayed in Fig.~\ref{ris:sp}. The characteristic energy of the accelerated electrons in the optimum near spherical laser bullet Eq.~(\ref{R_tau}) depends on the steady-state diameter, see Eq.~(\ref{W_maxS}). So for near spherical laser pulses of the same energy, a decrease of the cavity diameter results in the average electron energies decrease. This also happens for increased pulse length subject to self-modulation. From Tab.~\ref{tab:params_new} it is clearly seen that for a 40-fs laser pulse a decrease of $D$ leads to the drop of the average electron energies from 250 MeV ($0.005n_c$) to 95 MeV ($0.065n_c$). Here 
the average electron energy $\overline{\varepsilon}_{> 30 MeV}$, electron bunch charge $Q_{> 30 MeV}$ and conversion rate $\eta_{> 30 MeV}$ are calculated for the particles with the energies exceeding 30~MeV. 

\begin{table*}
\caption{Comparison of the accelerated electron bunch characteristics for the several sets of laser-plasma parameters and the same $W_L = 2.2$ J.}
\label{tab:params_new}    % Give a unique label
\begin{center}
\begin{tabular}{|c|c|c|c|c|c|c|c|c|}
\hline\noalign{\smallskip}
$P$, TW \quad &  $\tau$    &  $D_{L}$   &\quad $a_0$ \quad  & $n_e/n_c$  & \quad $\overline{\varepsilon}_{> 30 MeV}$ \quad & \quad $Q_{> 30 MeV}$ \quad & \quad $\eta_{> 30 MeV}$ \quad \\ \hline
\quad 210 \quad & \quad 10  fs \quad &\quad 2.8  $\mu$m \quad &  41.6  & \quad 0.15 \quad  & 150 MeV   &  10 nC  &  53\% \\
\quad 52 \quad & \quad 40  fs \quad & \quad 2.8  $\mu$m \quad & 20.8 &\quad 0.065 \quad & 95 MeV    & 6.7 nC   & 26\% \\
\quad 52 \quad & \quad 40  fs \quad & \quad 4.2  $\mu$m \quad & 13.9 &\quad 0.02 \quad & 140 MeV & 5.9 nC   & 35\% \\
\quad 52 \quad & \quad 40  fs \quad & \quad 5.5  $\mu$m \quad & 10.4  & \quad 0.005 \quad  &  250 MeV &  3.1 nC  & 35\%  \\

\noalign{\smallskip}\hline
\end{tabular}
\end{center}
\end{table*}

Unlike the average energy gain of accelerated electrons, the total charge of electron bunches decreases with $D$, although the conversion rate $\eta_{> 30 MeV}$ slowly increases. In general, it is seen that a bunch charge is higher for a shorter pulse (cf. the top and bottom rows in Tab.~\ref{tab:params_new}). However, for the 40 fs pulse, plasma density increase leads to the growth of the total electron bunch charge from 3~nC for $0.005n_c$ to 6.7~nC for $0.065n_c$ not displayed by the estimate Eq.~(\ref{Q_0}). This is because Eq.~(\ref{Q_0}) is derived for the RST condition Eq.~(\ref{RST}) and its accuracy is not enough to describe total charge increase for the mismatched conditions promoting self-modulation instability. It is worth noting that the further increase of the plasma density ($>0.065n_c$) in our simulations results in the total charge decrease with $n_e$. We attribute the latter to more significant laser pulse energy losses due to the laser self-modulation and confirm the existence of an optimum (over the accelerated charge) regime mismatched with RST (cf. \cite{Perevalov_2020}). As concerning comparison the results for the spherical "laser bullets" with different pulse durations (the top and bottom rows in Tab. \ref{tab:params_new}), the higher total charge of electron bunch is achieved for smaller spatio-temporal sizes of the laser pulse, that is in agreement with~Eq.~(\ref{Q_0}), $Q_0 \propto \tau^{-1/2}$.  However the charge ratio $Q_{10 fs}/Q_{40 fs}$ for the RST regime is also somewhat higher than predicted from rough estimate (3 instead of 2). Although such accuracy 33\%-50\% is quite reasonable for simple estimate,  one can assume that difference between numerical and theoretical results can be overcome with more accurate model accounting for the effect of already trapped in cavity electrons on the entire injection dynamics, the laser depletion length and the cavity gamma-factor \cite{Kostyukov_2009}. 

Conversion of the laser energy into the accelerated electrons energy is roughly proportional to the total number of accelerated particles and their averaged energy. The shortest and most tightly focused laser pulse for the RST conditions is beneficial to provide highest conversion rate that is clearly demonstrated by our simulations and estimates. For the 10~fs pulse a large value of $a_0$ makes it possible a strongly nonlinear regime of self-focusing in the form of laser bullet, which gives unprecedented conversion rate, 53\%. For longer pulses, which cannot ensure the absence of self-modulation at the initial stage of pulse propagation, the conversion rate is lower even though the RST mode is eventually established. For the relevant examples with a 40 fs laser pulse and a plasma with certain densities, higher  lateral transfer energy losses are realized for smaller cavities, which are formed in denser plasma targets. In a plasma with density $n_e=0.065n_c$ a conversion rate is 26\%, while for $n_e = 0.02n_c$  it was already 35\% (cf. second and third rows in Tab. \ref{tab:params_new}). On the other side, for larger cavity and lower density plasma, when laser pulse did not undergo self-modulation, a conversion rate stops to increase since laser field reaches its marginal magnitude for RST ($a_0 \sim 10$). 

Thus, for a given laser energy a pulse compression makes it possible to reach maximum both total charge of accelerated electron bunch and conversion rate. We also found that initially modulation unstable in rather dense plasma may evolve to the RST steady-state laser bullet with a high enough total bunch charge conversion efficiency. As a final study, in the next section, we refuse the condition of the same energy but consider the same laser focal spot to learn more on RST and electron energy gain. 

\section{Electron acceleration by pulses of different energies}{\label{sec6}}

\begin{figure*}
\center{\resizebox{0.85\hsize}{!}{\includegraphics{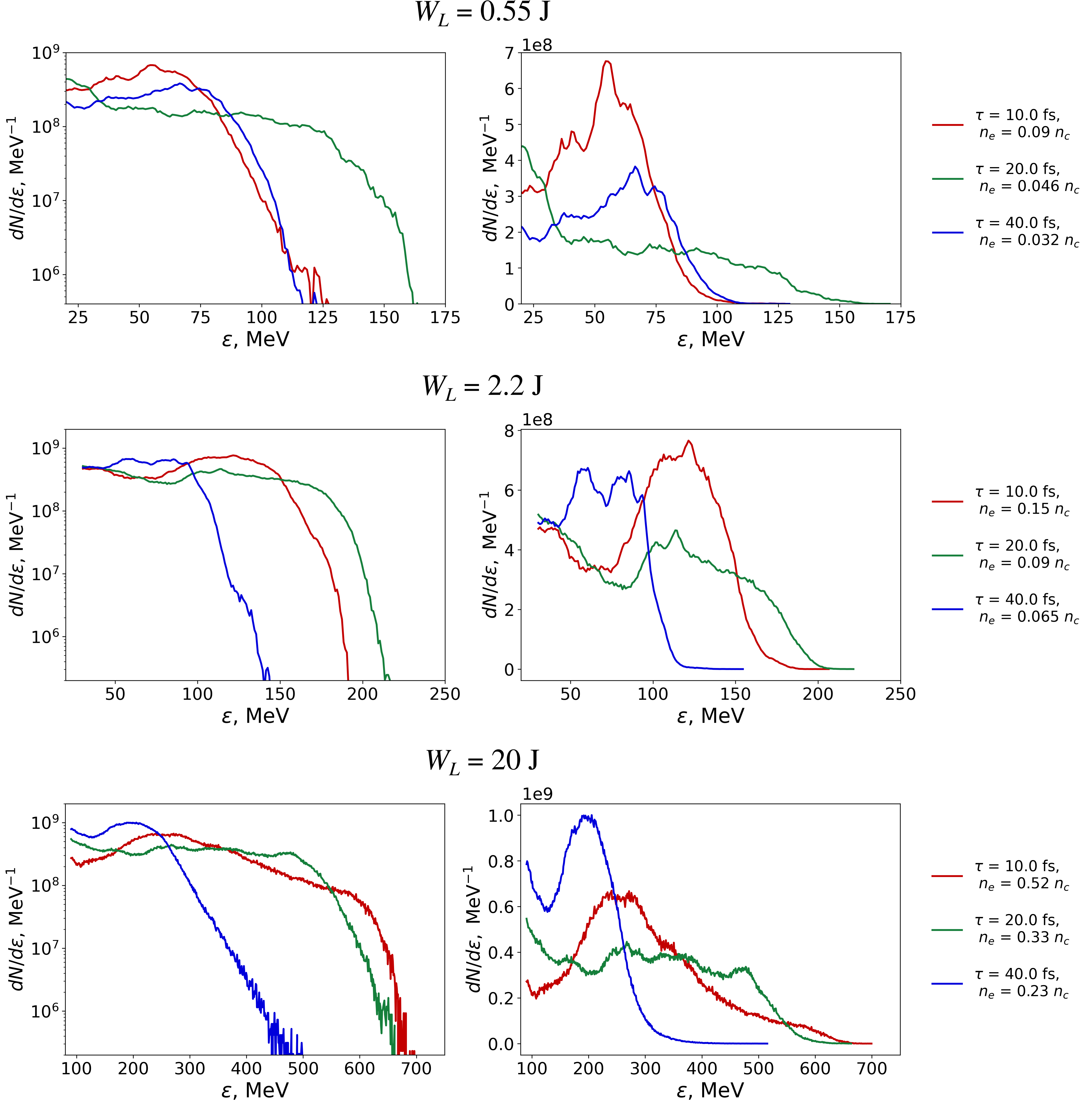}}}
\caption{Electron spectra at the instant when electron bunch charge becomes maximum for laser pulses with different $W_L$ focused in $ D_L=2.8\,\mu$m in the logarithmic (left) and linear (right) scales.}
\label{ris:sp1}
\end{figure*}

A series of simulations were performed for laser pulses with different energies and for the same focal spot size  (2.8~$\mu$m). The target densities were chosen to provide the plasma cavities with approximately the same diameters. Figure~\ref{ris:sp1} demonstrates the electron spectra for the parameters shown at right. It is seen how characteristic particle energy grows with laser pulse energy. 
Calculations of the total bunch charge and the conversion efficiency were done for the high-energy electrons with $\varepsilon>\varepsilon_{min}$ by using the energy cutoff, $\varepsilon_{min}$, corresponding to the beginning of a plateau in the electron spectra, approximately 15, 30 and 70~MeV for the laser energy of 0.55, 2.2 and 20~J, respectively. The table~\ref{tab:table2_new} summarizes the data for these charges and conversion rates. 

\begin{table*}
\caption{Laser-plasma parameters and corresponding total bunch charges and conversion rates for the laser focal spot $D_{L} = 2.8\mu$m.}
\label{tab:table2_new}    % Give a unique label
\begin{center}
\begin{tabular}{|c|c|c|c|c|c|c|}
\hline\noalign{\smallskip}
$W_L$, J \quad  &  $\tau$   & \quad $a_0$ \quad  & $n_e/n_c$ & \quad $Q_{> \varepsilon_{min}}$ \quad  & \quad $\eta_{> \varepsilon_{min}}$ \quad  \\ \hline
                 & \quad 10  fs \quad & 20.8 & 0.09 & 4.7 nC   &    50\%  \\
\quad 0.55 \quad & \quad 20  fs \quad & 14.7 & 0.046 & 3.4 nC & 40\%   \\
               & \quad 40  fs \quad & 10.3 & 0.032 &  3.2 nC    & 31\%  \\
\hline 
                & \quad 10  fs \quad  & 41.6 & 0.15 & 10.0 nC &  54\% \\
\quad 2.2 \quad & \quad 20  fs \quad   & 29.4 & 0.09 & 8.7 nC & 43\%  \\
             & \quad 40  fs \quad  & 20.8 & 0.065 & 6.7 nC & 26\% \\
\hline 
                & \quad 10  fs \quad  & 125 & 0.52 & 29 nC &  45\%  \\
\quad 20 \quad & \quad 20  fs \quad & 88.2 & 0.33 & 27 nC & 39\%  \\
 & \quad 40  fs \quad & 62.4 & 0.23 & 29 nC & 27\%\\
\noalign{\smallskip}\hline
\end{tabular}
\end{center}
\vspace*{-0.8cm} % with the correct table height
\end{table*}

For any laser energy considered, the maximum bunch charge is achieved for the shortest laser pulse. The same or even more pronounced is observed for the conversion rate. Note the good versatility on energy of the conversion rate value, which turns out to be approximately at the same level for the same durations. 

As a final point, we check the scalings discussed in Sec.~\ref{sec2}, i.e. whether the simulation data really follow the dependencies on the laser energy proposed there using the examples of 10~fs laser pulses.    
For the options discussed, the pulse energies were in the ratio $1:2^2:6^2$. The ratio of the total charges from simulations, $Q_{> \varepsilon_{min}}$, was $1:2:6$ (see fifth column of the Tab. \ref{tab:table2_new}). This well follows the square root dependence, Eq. (\ref{Q_0}), of the total charge on the laser energy.

The ratio of the average electron energies for the range $\overline{\varepsilon}_{> \varepsilon_{min}}$ was found from simulations as 74 MeV, 150 MeV and 415 MeV for 0.55, 2.2 and 20~J laser energies, correspondingly. This ratio reads $1:2:5.5$ and also corresponds well to the square root estimate Eq. (\ref{W_max}). Thus, we conclude that the scalings from Sec.~\ref{sec2} can be used for rough predictions of the electron bunch performance in experiments with various laser installations. Similar scalability of the laser-plasma accelerators has been also demonstrated for the bubble regime \cite{Gordienko_2005, Pukhov_2006}.

\section{Conclusion}{\label{sec7}}

We have demonstrated that to provide most efficient conversion of laser energy to electron bunch energy the pulse duration should be as short as possible. To ensure formation of the laser bullet, when the conversion rate is maximum, ultra relativistic intensities are required, $a_0 > 10$. This can be achieved by laser pulse compression, e.g. by using CafCA approach \cite{Khazanov_2019} and makes it possible for electron acceleration in the denser plasmas. In the denser medium the generated particle bunch gains the higher total charge, e.g. 3 nC and 10 nC for the laser pulses of the same energy, 2.2 J, and of 40 fs and 10 fs durations, correspondingly. In rather dense plasma the longer laser pulses undergo self-modulation that results in additional energy loss. However, there may be cases where the total accelerated charge can be higher than in the low dense plasma. 

It has been shown that the characteristics of the electron bunch are scalable with laser pulse energy and duration according to simple estimates, namely $Q_0 \propto \sqrt{W_L/\tau}$ and $\varepsilon_{max} \propto (W_L \tau)^{1/2}$. 
Whereas to provide the highest conversion rate and total charge electron bunch with reasonably high but not extreme energies the compressed laser pulse interacting with the high density plasma is preferable, the production of monoenergetic particle beam with the highest energy requires low density plasmas.    

The research performed could be of interest as a base for radiation-nuclear applications, such as betatron and Bremsstrahlung x-ray/gamma sources, photo-nuclear neutron and isotope production, meson factory, radiotherapy electron source, etc. Note, that high efficiency of the electron production with RST regime already opens the way for such applications with currently available commercial lasers.    

This work was supported in part by the Ministry of Science and Higher Education of the Russian Federation (agreement no. 075-15-2021-1361) and the Theoretical Physics and Mathematics Admancement Foundation ``BASIS'' (grant no. 22-1-3-28-1).

\newpage

\end{document}